# A Comparative Quantitative Analysis of Contemporary Big Data Clustering Algorithms for Market Segmentation in Hospitality Industry

Avishek Bose*, Arslan Munir†, and Neda Shabani‡

Email: *abose@ksu.edu, †amunir@ksu.edu, and ‡nshabani@ksu.edu

*Abstract*— The hospitality industry is one of the data-rich industries that receives huge Volumes of data streaming at high Velocity with considerably Variety, Veracity, and Variability. These properties make the data analysis in the hospitality industry a big data problem. Meeting the customers' expectations is a key factor in the hospitality industry to grasp the customers' loyalty. To achieve this goal, marketing professionals in this industry actively look for ways to utilize their data in the best possible manner and advance their data analytic solutions, such as identifying a unique market segmentation clustering and developing a recommendation system. In this paper, we present a comprehensive literature review of existing big data clustering algorithms and their advantages and disadvantages for various use cases. We implement the existing big data clustering algorithms and provide a quantitative comparison of the performance of different clustering algorithms for different scenarios. We also present our insights and recommendations regarding the suitability of different big data clustering algorithms for different use cases. These recommendations will be helpful for hoteliers in selecting the appropriate market segmentation clustering algorithm for different clustering datasets to improve the customer experience and maximize the hotel revenue.

## I. INTRODUCTION

IN recent years, the hospitality industry has emerged as one of the most profitable and vibrant businesses around the world. The hospitality Industry is perceived as the main source of revenue for many countries around the globe today. Many articles [4, 19, and 20] have shown that the growth of this industry will increase because of its growth. As the world steps into the Internet era with widespread utilization of Internet-connected appliances, the hospitality industry has transformed into a vastly data-rich industry [11, 12]. However, a structured way of utilizing available customer data for providing targeted recommendations [2] to customers is still missing. There are several other businesses like e-commerce websites and online stores that provide product recommendations [6, 7, 8, 13] to target potential customers. This trend of providing recommendations, such as customized offers and promotions, to customers via various mediums, such as websites, online social media, TV, and smart phones, is increasing day by day. However, it is infeasible to translate theses existing recommendation systems to the hospitality industry because of the vast scale of the hospitality network (i.e., customers, vendors, and proprietors) and its strict dependence on global economic trends. Furthermore, the hospitality industry requires an automated and dynamic recommendation system that renders many of the existing techniques focusing on offline recommendation systems ineffective.

In order to develop an effective customer recommendation solution for the hospitality industry, it is necessary to properly utilize the massive volumes of data gathered from customers. An effective recommendation system can help hoteliers to better meet customer preferences thus resulting in increased customer satisfaction as well as overall increase in hotel revenue. [1, 10] suggest that identifying market segmentation could be the key criterion to driving the hospitality industry forward in this regard. As technologies such as online social media, websites, smartphone, etc., become increasingly prevalent, it is imperative that the hospitality industry also utilize these platforms for providing recommendations, customized offers, and promotions to their customers. Market analysts have identified many aspects, goals, and processes involved in market segmentation for customer recommendation [3]. One of these processes is data clustering [9] which makes market segmentation expedient for market professionals.

Market segmentation for large data volumes can be carried out using big data clustering algorithms. Data clustering is a concept where similar kinds of points or objects of a dataset are grouped to remain in the same class. Thus, the points in the dataset are classified by their proximity to each other based on parameters given to the clustering algorithm. Although several clustering algorithms have been proposed in the literature, there is little to no information available as to the suitability of one algorithm over another with regards to big data clustering in the hospitality industry. As hospitality datasets are significantly heavy featured, a factual review is a must for making an informed choice on the appropriate clustering algorithm.

There exist various types of clustering algorithms, namely: (i) centroid-based clustering, (ii) hierarchical clustering, (iii) distribution-based clustering, (iv) density-based clustering, and (v) grid-based clustering.

In particular, several papers have discussed efficient density-based algorithms, such as DBSCAN [14], OPTICS [15], EnDBSCN [16], and few other variation of these algorithms [5, 17], however, each of the algorithm has its limitations and weaknesses. We restrict our analysis to density-based algorithms because market segmentation using these algorithms can be done efficiently. Furthermore, density-based algorithms incorporate various significant factors of clusterization, such as the number of actual noise points, number of actual clusters, etc., in the datasets. In general, the efficiency of a big data clustering algorithm is contingent on how many input parameters the algorithm depends on and its clustering performance in different scenarios, such as varying densities, embedded clusters, and nested adjacent clusters.

DBSCAN is known as the first authentic density-based clustering algorithm. However, DBSCAN does not provide accurate results to identify clusters of varying densities as well as embedded or adjacent clusters. Because of augmented ordering of data-points, OPTICS requires the overhead of calculation, and it also faces some problems in identifying embedded or nested clusters. Both DBSCAN and OPTICS need efficient input parameter setup for getting the desired clustering from the given datasets. Similarly, EnDBSCAN has two problems: the first one is repeated-analysis of data points in boundary lines within a cluster and the second one is inefficient clustering for nested adjacent clusters. Two recent research approaches try to overcome the limitations of DBSCAN, OPTICS, and EnDBSCAN. The first one formulates ascendingly sorted k-distance graph of first order derivative which incurs additional calculations, and the second one requires three initial parameters, which directly indicates that this approach will be dependent on those parameters.

Our main contributions in this paper are:

- We have presented a detailed review of various clustering algorithms and classified them based on their usefulness for market segmentation in the hospitality industry for various use cases.
- We have characterized the limitations, performance, complexity, and usefulness of various clustering algorithms for different use cases.
- We have implemented various density-based clustering algorithms, such as DBSCAN, OPTICS, and EnDBSCAN, and have provided a comparative performance analysis of these clustering algorithms for different data sets.
- Based on our analysis and implementation, we have characterized requirements of developing future clustering algorithms for market segmentation in hospitality industry.

The rest of this paper is organized as follows: Section II discusses the motivation for this work. Section III presents the background study. The literature review is presented in Section IV. Section V presents simulation results and performance analayis. Finally, Section VI concludes the work and identifies future research directions.

## II. MOTIVATION

The hospitality industry is heavily dependent upon the Internet and electronic transactions (e.g., online bookings, point-of-sale transactions, etc.). A report [20, 21] published a few years ago noted that 52.3% of all hotels and other bookings related to the hospitality industry had been made online in 2010. This trend is still going upward. For generating an effective recommendation for the customer, an effective clustering algorithm is needed to address the challenges discussed above in Section I. Although many clustering algorithms (e.g., algorithms based on the density of point) exist, most of these algorithms have a problem in identifying clusters of varying densities and embedded clusters. Fig. 1 shows an embedded cluster and Fig. 2 shows a cluster of varying density.

The customer data in hospitality industry is likely to contain embedded clusters and clusters of varying densities. As an example of varying density feature, consider a hospitality dataset indicating that a majority of the U.S. citizens in all age ranges visit a sea beach at least once a year. Specifically, U.S. teenagers visit beaches more frequently than people in other age ranges. If we have a dataset of visitors based on age and number of visits, we can apply clustering algorithm over that dataset. From the dataset, we might observe that the region belonging to teenager citizens is denser than the other regions based on the number of datapoints or times of visits to the beach. This variation of dense regions represents the varying density property of clusters.

As an example of a nested embedded cluster, consider a hospitality dataset related to U.S. citizens' visits to Europe. The data shows that a majority of the U.S. citizens visit Europe. Specifically, people residing in the East Coast visit British Isles frequency, and the people residing in the West Coast visit Spain frequently. Furthermore, the U.S. citizen with age range in between fifty to seventy years and living in East Coast usually visit historical British monarch's places. If hoteliers collect and cluster the dataset of U.S. citizens based on the citizens' place of residence and age, the U.S. citizens of older age and residing in the East Coast should be in the core cluster of an embedded cluster as a potential visitor to London. The people who reside in the East Coast but are not old are likely to be potential visitors of the British Isles and their cluster will be the outer cluster encompassing the core cluster of London visitors. The outermost cluster will consist of all U.S. citizens who visit Europe.

Hence, in order to evaluate clustering algorithms utilized in hospitality industry, criteria, such as varying density, nested embedded cluster, etc., need to be considered.

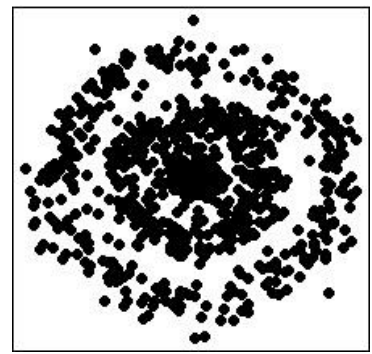

Fig. 1. Embedded cluster

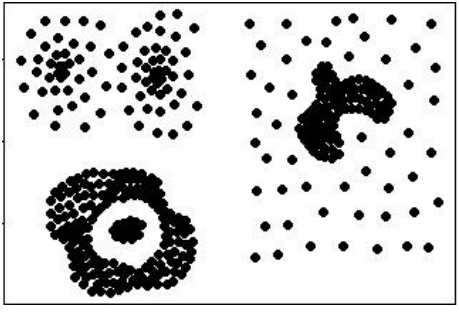

Fig. 2. Varying density cluster

There exist some research [5, 14, 15, 16, 17] targeting clustering algorithms, however, most of these algorithms have limitations. Majority of these algorithms are vastly dependent on user-defined parameters, and if those parameters are not properly selected, significant changes in results can occur. Furthermore, the complexity of these algorithms is also a matter of huge concern because if it is not addressed properly, *dynamic market segmentation* would not be possible, which could impact the hospitality industry business. *Dynamic market segmentation* is an automated process to generate recommendations for the customers at runtime using clustering. To address the limitations of existing clustering algorithms, a detailed analysis of these clustering algorithms is imperative.

## III. BACKGROUND

In this section, we have presented basic definitions and ideas related to density-based clustering algorithms.

*Definition 3.1:* Density-based clustering works by differentiating the density of points in a specific area. For example, the density of one area could be higher than the density of another area based on the number of points present in the specified or given area. Let $p$ is the point in the dataset $D$, the density of a specified point $p$ is measured by the number of data-points $|N_{\mu(p)}|$ present in $p$'s neighborhood $\mu$. $|N|$ denotes the number of points or objects in the neighborhood of any specific point or object.

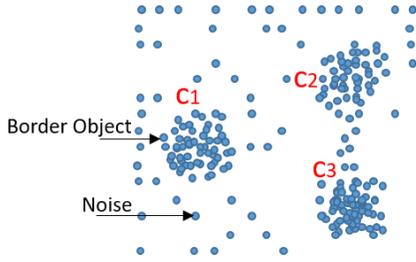

Fig. 3. Illustration of three clusters($C_1$, $C_2$, $C_3$) with noise and border-point

*Definition 3.2:* Neighborhood $\mu$ of a point $p$ is considered a circular area generated by a given parameter radius $r$ as an input value, centering at the point $p$. If any point $q$ from the dataset $D$ is in the circular area of $p$ and their shortest distance is $dist(p,q) \leq r$, it can be said that $q$ is in the neighborhood of $p$ or in other words point $q$ is point $p$'s neighbor. So, neighborhood of $p$ can be defined as $\mu_p \rightarrow \{q \in D \,/\, dist(p,q) \leq r\}$

*Definition 3.3:* The number of points that must be present in the neighborhood of a point $p$ to make it as a core-point to form a cluster is referred as *MinPts*. The number, size as well as shape of a cluster is heavily dependent upon this user given parameter. Additionally, except for the border point, a neighborhood of a particular point inside a higher density cluster has more data-points than *MinPts*, but the points inside lower density cluster may have at least the equal number of points as *Minpts*. As *MinPts* can only be a natural number, therefore *MinPts* $\in \mathbb{N}$ where $\mathbb{N}$ denotes the set of natural numbers.

*Definition 3.4:* Core-point or core-object $p$ of a cluster $C_k$ (where $k = 1, 2, 3,...,n$) are those data-points in the cluster $C_k$ which have equal or greater number of points as (*MinPts*) in its neighborhood $\mu$. Core-object $\theta \rightarrow \{p \in C_k \,/\, |N_{\mu(p)}| \geq MinPts\}$ where $\theta$ refers to the set of all core-points.

*Definition 3.5:* Border-point or border-object $s$ of a cluster $C_k$ (where $k = 1, 2, 3,...,n$) is that data-point in the cluster $C_k$ which do not have a sufficient number of points as (*MinPts*) in its neighborhood $\mu$, but still those are the member of that cluster. Border point $\lambda \rightarrow \{e \in C_k \,/\, |N_{\mu(e)}| < MinPts\}$ where $\lambda$ refers to the set of all Border point.

*Definition 3.6:* Noise points are that data-points which are not the member of any cluster. The area contains noise points has a low density of points than the other areas that contain clusters. In another way, if any point except the border-point in the dataset doesn't have an equal number of points as *MinPts* in its neighborhood, this point can be referred as noise. Let the dataset $D$ has $n$ number of clusters represented by the cluster set $Z_c = \{C_1, C_2...C_k...C_n\}$ where $k, n \in \mathbb{N}$. If $\omega$ is noise point, then noise $\omega \rightarrow \{\omega \in D \,/\, \forall_{k=\omega} \notin Z_c\}$, where $\omega$ refers to the set of all noise points. Fig. 3 represents three clusters named $C_1$, $C_2$ and $C_3$ as well as noise with border-point.

Definition 3.7: Core-distance $\gamma$ of a point $p$ is the minimum distance of neighborhood of the point which contains an equal number of points as *(MinPts)* inside its neighborhood. Core distance $(\gamma) \rightarrow \{/\gamma/ \leq r_\mu, |N\mu_{\gamma(p)}| = MinPts\}$.

Here $r_\mu$ is the given radius of the neighborhood, $\mu_{\gamma(p)}$ is the neighborhood covering $p$'s core-distance and $|N\mu_{\gamma(p)}|$ is the number of point of that neighborhood.

Definition 3.8: Let $p$ is a core point or object, and $q$ is another point or object in the dataset. Reachability-distance $\phi$ of the object $q$ is the shortest distance from $p$ if $q$ is directly density-reachable from $p$. Reachability distance $\phi(q,p) \rightarrow \{q \in /N_{\mu(p)}|, |r_{p(r)}| \geq \phi_q \geq \gamma_p\}$.

Reachability-distance $\phi$ cannot be smaller than the core-distance $\gamma$. Fig. 4 illustrates core-point $p$, a point $q$ in the neighborhood of $p$, core-distance $\gamma_p$ of $p$, and reachability distance $\phi(q,p)$ of $q$ from the point $p$ to $q$.

Definition 3.9: A point $q$ can be said directly density-reachable from a point $p$ if $q$ is located inside the $p$'s neighborhood $\mu(p)$ and point p has at least number of points equal to *MinPts*.

Definition 3.10: A point *q* can be referred as density-reachable from a point *p* if those are connected via directly density-reachable points. Let a chain of points is $p_1, p_2, p_3 ... p_n$ and any point $p_{j+1}$ of this chain is directly density-reachable from $p_j$ where j $\in$ {*1, 2, ...,n-1*}. If $p_1=p$ and $p_n=q$, *q* is density-reachable from *p*.

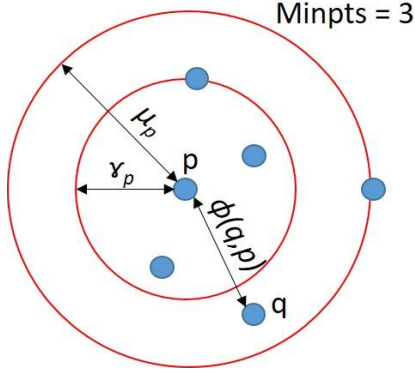

Fig. 4. Illustration of core-object *p*, core-distance $r_p$, and reachability-distance $\phi(q,p)$

Definition 3.11: A point *q* can be called density connected to a point *p* if both *p* and *q* are density-reachable from another point *o*.

Fig. 5 illustrates the graphical representation of directly density-reachability, density-reachability, and density-connectivity. In this figure, both points *q* and *r* are directly density-reachable from the point *p* whereas *q* and *r* are density-reachable via point *p*. The points t and s are density-connected through the density-reachable points *p*, *q*, and *r*.

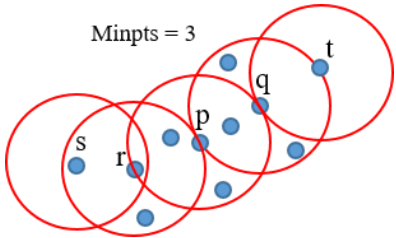

Fig. 5. Illustration of Directly Density-reachable, Density-reachable and Density-connected Points

## IV. LITERATURE REVIEW

### A. DBSCAN:

DBSCAN [14] is one of the significant early approaches in density-based methods to cluster points of a dataset. It works by traversing all the points of a dataset, and picks a point arbitrarily. If *p* is an arbitrary point selected from a dataset, this algorithm is able to access all the points within the point *p*'s neighborhood $\mu$. If *p* is a core object, it can access all neighborhood points and can let the process repeat for its neighbor points to expand a cluster, but this is not true for the border point $\lambda$. While any border point $\lambda$ is chosen arbitrarily to access its neighborhood, this algorithm skips that point because it won't satisfy the condition to access the neighbor points due to less number of points than *MinPts* of its neighborhood. The current cluster id is assigned at that border point and starts accessing next arbitrary point. If the chosen point is a noise, it will not satisfy density-connectivity feature of this algorithm. DBSCAN faces some difficulties to identify a varying density space because it uses global neighborhood radius $r_\mu$ and *MinPts*. That's why it cannot perform well to detect varying density cluster and embedded cluster. If the two clusters are in close proximity or are adjacent to each other, the process may detect those as a single cluster. The same thing may happen if clusters of varying density are located one inside another such as a nested embedded cluster. In the case of detecting nested embedded cluster, outcomes go beyond the performance of this approach. If two adjacent clusters do not have more distance than given neighborhood radius $r_\mu$, it is not possible to make any distinction between two sets of points to identify those as two separate clusters.

*Algorithmic Analysis:*

The runtime complexity of DBSCAN algorithm for each point is the runtime needed for the query to process all the neighbor-points in the neighborhood $\mu$. As this process would be performed for each object of the datasets, the optimal runtime for DBSCAN algorithm is $O(n \ log_k n)$ where *n* is the number of object in the datasets and *k* is the number of the core-object. The optimal runtime complexity is only applicable if tree based spatial index is used otherwise the complexity could be $O(n^2)$.

### B. OPTICS:

OPTICS [15] is another clustering algorithm based on density analysis which orders points by comparing point's reachability-distance to the closest core-point that is directly density-reachable from those points to identify cluster. However, this algorithm does not directly identify cluster from the dataset because after ordering of the objects, any density-based clustering approach such as DBSCAN does rest of the task of clustering. According to the procedure of OPTICS algorithm after creating an augmented ordering of cluster points, this approach can be used with any other density-based approaches such as DBSCAN [10]. The ordering stores core-distance and suitable reachability distance for each point. After formulating reachability plot, an optimum neighborhood radius $r_{opt}$ might be selected to generate the proper result of clustering.

*Algorithmic Analysis:*

OPTICS is evolved to encounter the limitation of DBSCAN [14] such as to detect the varying density of cluster objects. It provides a handy solution to meet the problems of global density parameter issue and varying density efficiently. As DBSCAN was vastly dependent upon input parameters like neighborhood radius $r_\mu$ and *Minpts*, ordering of objects has minimized the dependency for those parameters in OPTICS. Although this algorithm elaborately discusses visual techniques of cluster ordering, reachability plots, etc. to counter the dependency of input parameter, actually visual technique also need some parameter setting such as the threshold value of neighborhood or optimum neighborhood radius $r_{opt}$ to identify clusters. Afterwards, the process of selecting threshold value can be crucial to identify cluster. If an inappropriate threshold value is selected, some clusters are likely to be undetected, and the algorithm will not be able to detect embedded clusters. Moreover, this process experimented using the specific datasets to get range values, but whether these values are feasible or not for all datasets like hospitality datasets, is not specifically mentioned in this approach.

As OPTICS requires ordering of points as an extra calculation, its complexity is higher than other density based algorithm. If it uses any tree based spatial index, its runtime would be $O(n \log n)$ otherwise it would be $O(n^2)$. Only if the algorithm has direct access to the neighborhood $\mu$ or organized in a grid, the runtime requires to cluster from the ordered-dataset is $O(n)$. So, the overall runtime complexity of OPTICS for extracting the clusters from the datasets is at least $O(n \log n) + O(n)$.

*C. EnDBSCAN:*

The main idea of EnDBSCAN [16] algorithm is that if the difference of the core distance between two points is in the range of a pre-defined variance factor, both the points are identified to be in the same cluster. EnDBSCAN also starts clustering by selecting an arbitrary point from a dataset and calculates its core-distance considering the given parameters *MinPts* and neighborhood radius $r_\mu$. If the point's core-distance is greater than given neighborhood radius $r_\mu$, it is considered as noise point. When core-distance is smaller or equal to the given neighborhood radius $r_\mu$, the point is considered as a core-point. Then the core-point is allowed to expand the cluster through its neighborhood points within the range of its core-distance $\gamma$. After assigning a new cluster id to the core point or object, all the core-neighbors of this point are assigned the same cluster id. The process of expanding and clustering repeats until all the dataset's points have been assessed. To remain in the same cluster, the difference between the core-distance of an initially selected arbitrary point and the core-distances of core-neighbor points of that arbitrary point cannot be more than a predefined parameter $\beta$. If the difference does not satisfy this condition, it indicates a density variation between points and the points must be in different clusters. However, this situation occurs only in the boundary region of two different clusters, and requires repetition of this process for border points located in the border region of two different dense area.

*Algorithmic Analysis:*

If a spatial index tree is used, the runtime complexity of EnDBSCAN will be $O(n \log n)$ like DBSCAN. If there are many clusters in a dataset such as those in hospitality datasets, processing runtime complexity of repetitive border-points need to be taken into consideration. However, if there are only a few number of clusters within a dataset, the runtime complexity of process repetition for border-points can be neglected.

*D. A variant of DBSCAN Algorithm to Find Embedded and Nested Adjacent Cluster:*

A variant of DBSCAN algorithm has been proposed in [17] to counter the limitation of previously presented density-based algorithms. To estimate the value of neighborhood radius $r_\mu$ as an input parameter, it uses the concept of *k*-distance plot and first-order derivative instead of selecting them by datasets observation. This approach allows the user to input the value of

*MinPts*. To expand the cluster, firstly an arbitrary point has to be checked to verify the possibility of being a core-point. If the selected point is a core-point, only then the expansion process of clustering can be performed. Furthermore, this approach introduces a new term named neighborhood-difference. The term neighborhood-difference is defined as the difference between the numbers of neighborhood points of those two points. For example, one point belongs to a cluster as a core-point and another point is in the former's neighborhood with respect *MinPts* and neighborhood radius $r_\mu$, to determine those points are in the same cluster or not, the value of neighborhood-difference of those points must be within the range of tolerance factor $\alpha$. The tolerance factor $\alpha$ is a value given as an input parameter by the user. If the neighborhood difference of that two points is greater than the tolerance factor $\alpha$, those points might not be in the same cluster. Instead of expanding cluster through neighborhood $\mu$ expansion like DBSCAN, this approach expands through core-neighborhood $\mu_\gamma$ of a cluster by satisfying the tolerance factor $\alpha$ issue as discussed earlier. Then a sorted *k*-distance graph is formulated in a plot to get the effective value of neighborhood radius $\mu_r$. In this plot, a total number of points in the datasets take the independent (X) axis, and corresponding distances from each point to its $k^{th}$-nearest neighbor take the dependent (Y) axis. After sorting and completing the *k*-distance vector and the first-order derivation respectively, we can get the effective value of neighborhood radius $\mu_r$. If we see huge change of slope or sudden variation in the sorted *k*-distance graph, we can detect separation of cluster points from the noise points. Thus, we can also identify noise points by analyzing the threshold point from sorted *k*-distance graph. While sorting, if more than one data-points have equal $k^{th}$ nearest distance, it is also possible that a neighborhood can contain more than $k+1$ data-points.

*Algorithmic Analysis:*

The manuscript where this approach has been presented does not clearly mention runtime complexity of this approach. Because the algorithm requires to implement a *k*-distance graph, the runtime complexity of this process will be $O(n)$. The graph-vector needs to be sorted, and the complexity of this process will be at least $O(n \log n)$ if an efficient sorting algorithm has been applied. Furthermore, optimal runtime complexity considering the expansion of cluster will be $O(n \log n)$ if spatial index tree used otherwise it will be $O(n^2)$. So the total optimal complexity of this approach is $O(n) + O(n \log n) + O(n \log n)$.

*E. Effective Density-based Approach to detect Complex Data Clusters:*

Nagaraju et al. [5] have proposed a density-based approach to identify clusters of varying densities and nested adjacent clusters. This approach recognizes that variation in the neighborhood data-point is useful to identify cluster rather than cluster density variation. To address their analysis this approach defined a new term named tolerance factor $\delta$ which is an input value given by the user. According to this approach, difference in the number of core-neighbors of a specific core-point and the number of core-neighbors of that core-point's core-neighbors might be less or equal to tolerance factor $\delta$ to remain in same class. If the difference is more than the tolerance factor, this algorithm may detect it as noise point or object.

*Algorithmic Analysis:*

Although this approach is presented to minimize the dependency of global density parameter for clustering, this algorithm also requires efficient parameter setting such as the neighborhood radius $r_\mu$ and tolerance factor $\delta$. This algorithm also requires continuous adjustment of tolerance factor $\delta$ to identify cluster's border-points properly. The significant problem with this approach is that it may identify many insignificant clusters. As this algorithm calculates neighborhood-difference and no predefined number of *MinPts* is mentioned that may consist in a given neighborhood $\mu$, it might misleadingly identify too many clusters in the datasets. Whenever it finds the difference of neighborhood points, it may identify a new cluster.

As this approach hasn't specifically mentioned any use of spatial index tree, the runtime complexity of this algorithm will be O($n^2$).

## V. COMPARISON OF EXISTING CLUSTERING ALGORITHMS PERFORMANCE

Data related to human behavior and e-commerce is varied and complex. Thus providing recommendations by segmenting complex datasets such as hospitality industry datasets, requires efficient clustering algorithms which can identify varied density clusters and nested embedded cluster. Moreover, clustering algorithms should not be time-consuming if they are to be used for automated recommendation systems. The automated recommendation system is a kind of system which can generate a recommendation for the customer dynamically. Consequently, customer interaction with the system is also analyzed in real-time by the system to provide further effective recommendations. Therefore, runtime complexity is another criterion for measuring the performance of clustering algorithms.

As hospitality industry datasets are not coherent like medical imaging, animal genetic information datasets, and not predictable like e-commerce business market datasets, cluster analysis of these kinds of datasets is different. Scenarios such as varying density, nested adjacency, and nested embedded features of the cluster are very common in this industry's datasets. To address these scenarios properly in our study, we have used synthetic data. This approach produces significant results of clustering that help to evaluate the performances of algorithms mentioned in Section IV for hospitality big data.

In this section, we have first presented synthetic data relevant to the scenario mentioned above and then experimented with clustering algorithms over those datasets. We have performed algorithmic analysis of various density-based algorithms. Table 1 summarizes the runtime complexity of the implemented density-based algorithms.

TABLE I
OPTIMAL RUNTIME COMPLEXITY OF DISCUSSED DENSITY-BASED
CLUSTERING ALGORITHM

| Algorithms | Optimal Runtime Complexity |
|---|---|
| DBSCAN [14] | $O(n \log n)$ |
| OPTICS [15] | $O(n \log n) + O(n)$ |
| EnDBSCAN [16] | $O(n \log n)$ |
| DBSCAN variant [17] | $O(n) + O(n \log n) + O(n \log n)$ |
| Nagaraju et al. Approach [5] | $O(n^2)$ |

Fig. 6 presents the comparative results of clustering for the algorithms mentioned above using different synthetic datasets. We have collected these synthetic datasets from [18], which are relevant in the hospitality industry context. We have further used R compiler to refine and generate new datasets. Fig. 6(a) represents a generic dataset without varying density cluster property.Fig. 6(e) represents a varying density and nested cluster feature dataset, and Fig.6(i) represents a nested embedded cluster.

Firstly, Fig.6(b), Fig.6(c), Fig.6(d) show the results of DBSCAN, OPTICS, and EnDBSCAN, respectively for the dataset shown in Fig.6(a). Here, all the three density-based algorithms (i.e., DBSCAN, OPTICS, and EnDBSCAN) perform well to identify those clusters. Secondly, Fig.6(f), Fig.6(g), Fig.6(h) show the results of DBSCAN, OPTICS, and EnDBSCAN, respectively, for the dataset shown in Fig.6(e). For this dataset both OPTICS and EnDBSCAN perform well to identify those clusters whereas DBSCAN fails to identify some clusters because of varying density of points in the dataset. Finally, Fig.6(j), Fig.6(k), Fig.6(l) show the results of DBSCAN, OPTICS, and EnDBSCAN, respectively, for the dataset shown in Fig.6(i). For this dataset, only EnDBSCAN performs well to identify the nested embedded clusters. On the other hand, DBSCAN and OPTICS both fail to identify embedded cluster. OPTICS indentifies many insignificant clusters instead of identifying these as a single cluster, and DBSCAN cannot identify that this dataset consist of different clusters. Clustering performance of another density-based algorithm, a variant of DBSCAN algorithm [17], may be better than the original DBSCAN algorithm because the variant detects neighborhood radius and threshold point of noise from first order derivative of the *k*-distance graph. However, comparable clustering performance can also be achieved by using the OPTICS algorithm [15] if an optimum neighborhood radius $r_{opt}$ is selected from the reachability plot. The density-based approach [5] mentioned in Section IV may identify many insignificant clusters instead of identifying the correct cluster. Furthermore, the approach [5] also has some dependency on its input parameter such as tolerance factor $\delta$.

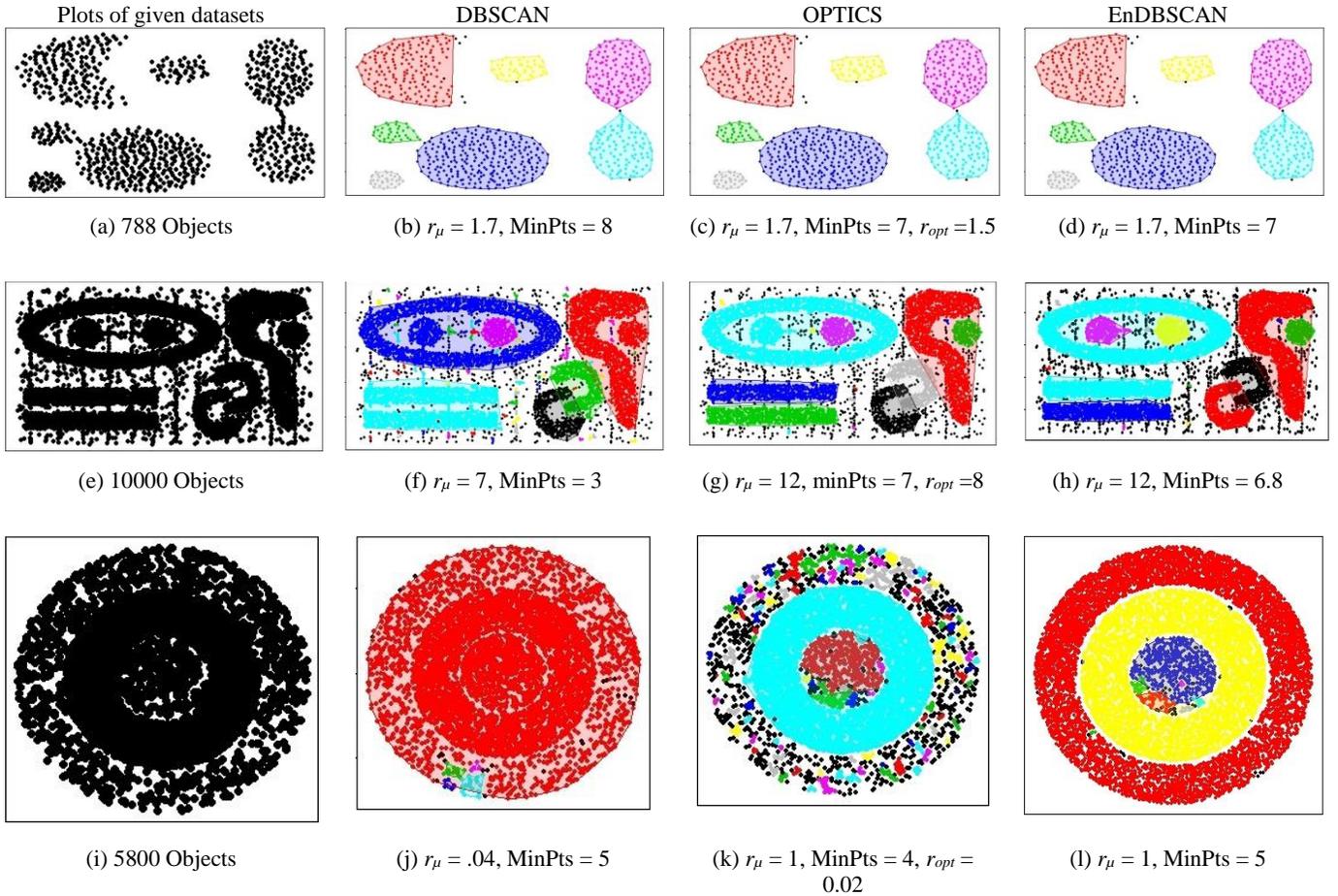

Fig. 6. Performance result of different density-based clustering algorithms using synthetic datasets relevant to the datasets of hospitality Big Data

## VI. CONCLUSION

Although hospitality industry is one of the leading business in the world and also increasing its economy steadily, very few research works have been conducted regarding the proper utilization of huge volume of available customer data. This paper provides insights into data clustering features of hospitality big data by analyzing existing density-based clustering algorithms. We have implemented popular density-based algorithms, such as DBSCAN, OPTICS, EnDBSCAN, and a few other variants of density-based algorithms, and have provided a comparative performance analysis of these algorithms. Results reveal that EnDBSCAN performs superior than DBSCAN and OPTICS in terms of identifying nested and embedded clusters. Similarly, OPTICS perform better than DBSCAN in identifying adjacent nested cluster for different datasets. However, all of the contemporary clustering algorithms have their limitations in identifying clusters from datasets because of their dependency on input parameters.

We can conclude that further research is needed to counter the limitations of existing clustering algorithms. Furthermore, novel clustering algorithms need to be developed for enabling automated recommendation systems for the hospitality industry to improve both customers experience and revenue of the hospitality industry.